\documentclass[comment, prl,twocolumn,nofootinbib, preprintnumbers, superscriptaddress]{revtex4-2}

\usepackage{booktabs}
\usepackage{multirow}
\usepackage{amsmath,amssymb,slashed,braket}
\usepackage{graphicx}
\usepackage{cancel}
\usepackage{epstopdf}
\usepackage{float}
\usepackage[colorlinks=true,
            linkcolor=blue,
            urlcolor=blue,
            citecolor=green,          
            bookmarks=true,
            bookmarksnumbered=true,
            breaklinks=true,
            pdfpagemode=Fullscreen,
            pdfstartview=FitBH]{hyperref}

\usepackage[normalem]{ulem}
\usepackage{color}

\definecolor{Orange}{cmyk}{0,0.61,0.87,0}
\definecolor{JungleGreen}{cmyk}{0.99,0,0.52,0}
\definecolor{OliveGreen}{cmyk}{0.64,0,0.95,0.40}
\definecolor{Brown}{cmyk}{0,0.81,1,0.60}
\definecolor{RoyalBlue}{cmyk}{0.71,0.53,0,0.12}
\definecolor{Gray}{cmyk}{0,0,0,0.40}
\definecolor{LightPink}{cmyk}{0.0,0.25,0,0}
\definecolor{LLightPink}{cmyk}{0.0,0.10,0,0}
\definecolor{LightBlue}{cmyk}{0.25,0,0,0}
\definecolor{LightGray}{cmyk}{0,0,0,0.2}

{}

\usepackage{xcolor}
\definecolor{gesfpurple}{rgb}{0.47,0.19,0.42}

\definecolor{gesflanse}{rgb}{0.00,0.50,0.50}

\definecolor{gesfblue}{rgb}{0.08,0.42,0.76}

\definecolor{gesfred}{rgb}{1,0,0}

\definecolor{gesfwhite}{rgb}{1,1,1}

\definecolor{gesfblack}{rgb}{0,0,0}

\newcommand{\geqn}[1]{Eq.\,\hypersetup{linkcolor=blue}(\ref{#1})\hypersetup{linkcolor=blue}}
\newcommand{\gfig}[1]{{\hypersetup{linkcolor=violet}Fig.\,\ref{#1}\hypersetup{linkcolor=blue}}}
\newcommand{\gtab}[1]{{\hypersetup{linkcolor=gesflanse}Tab.\,\ref{#1}\hypersetup{linkcolor=blue}}}

\graphicspath{{Figures/}}

\begin{document}

\title{
CP Prediction from Residual $\mathbb Z_2^s$ and $\overline{\mathbb Z}_2^s$ Symmetries with JUNO First Data}

\author{Shao-Feng Ge}
\email{gesf@sjtu.edu.cn}
\affiliation{Tsung-Dao Lee Institute \& School of Physics and Astronomy, Shanghai Jiao Tong University, China}
\affiliation{Key Laboratory for Particle Astrophysics and Cosmology (MOE) 
\& Shanghai Key Laboratory for Particle Physics and Cosmology, Shanghai Jiao Tong University, Shanghai 200240, China}

\author{Chui-Fan Kong}
\email{kongcf@ibs.re.kr}
\affiliation{Particle Theory and Cosmology Group (PTC), Center for Theoretical Physics of the Universe (CTPU),
Institute for Basic Science, Daejeon 34126, Republic of Korea}

\author{João Paulo Pinheiro}
\email{joaopaulo.pinheiro@fqa.ub.edu}
\affiliation{Tsung-Dao Lee Institute \& School of Physics and Astronomy, Shanghai Jiao Tong University, China}
\affiliation{Key Laboratory for Particle Astrophysics and Cosmology (MOE) 
\& Shanghai Key Laboratory for Particle Physics and Cosmology, Shanghai Jiao Tong University, Shanghai 200240, China}

\begin{abstract}
The JUNO first data and the recent neutrino global fit results
are implemented in the sum rule from the residual $\mathbb Z^s_2$
and $\overline{\mathbb Z}^s_2$ symmetries to make prediction
of the leptonic Dirac CP phase $\delta_D$. Without involving
model parameters, the probability distribution of $\delta_D$
can be readily obtained from the experimental measurements
of the three mixing angles. We then confront the theoretical
predictions with the global fit results for the CP phase
as well as the T2K and NOvA joint analysis for their CP
measurement to give the data preference of the two
residual symmetries with Bayes factor for both normal
and inverted orderings. We further extend our analysis
to a two-dimensional probability distribution to fully explore
the correlation between the CP phase $\delta_D$ and the
atmospheric angle $\theta_a \equiv \theta_{23}$.
\end{abstract}

\maketitle

%%%%%%%%%%%%%%%%%%%%%%%%%%%%%%%%%%%%%%%%%%%%%%%%%%%%%%%%%% 
\textbf{Introduction} -- 
% \gred{Para 1: Scientific Importance of Neutrino Mixing and CP}
Neutrino oscillation is the first new physics beyond the
Standard Model of particle physics with various solid
experimental data \cite{PDG24-Nu}. It provides not just a key
to understand the possible new physics world beyond our
current knowledge
\cite{SajjadAthar:2021prg,Arguelles:2022tki,deGouvea:2022gut,Huber:2022lpm},
but also a first observation of quantum
interference phenomena at macroscopic lengths spanning from
$\mathcal O(1)$\,km at reactor experiments such as Daya Bay
to $\mathcal O(10^5)$\,km for solar neutrino
transition\footnote{Note that the 2025 Nobel Prize in Physics awards
quantum phenomena at the centimeter scale.}. With the Daya Bay
\cite{DayaBay:2012fng}, RENO \cite{RENO:2012mkc}, and
Double Chooz \cite{DoubleChooz:2011ymz}
experiments establishing a nonzero $\theta_r (\equiv \theta_{13})$
mixing angle, neutrino physics has entered a precision era
\cite{Fogli:2012ua}.
Especially, this allows possible measurement of the
leptonic Dirac CP phase $\delta_D$
\cite{Branco:2011zb,Machado:2013kya} that may hold the key
to understanding why our Universe is made of matter but
almost no anti-matter \cite{Davidson:2008bu,DiBari:2012fz}.

% \gred{Para 2: Flavor Symmetries}
The mixing parameters will be measured with high
precision at the new-generation neutrino experiments.
Especially, the JUNO experiment can provide sub-percentage
precision for $\theta_s (\equiv \theta_{12})$ and
$\Delta m^2_s$ $(\equiv \Delta m^2_{21})$ 
\cite{Ge:2012wj,JUNO:2015zny} with full data collection.
This would allow further test of those flavor symmetries
and models that can predict the neutrino mixing pattern.

% \gred{Para 3: Residual Symmetry \& Sum Rule:}
A flavor symmetry is typically imposed on the fundamental
Lagrangian of an ultra-violet (UV) complete
theory as starting point
\cite{Altarelli:2012ss,King:2013eh,King:2017guk,Petcov:2017ggy,
Xing:2020ijf,Feruglio:2019ybq,Ding:2023htn}.
Being an UV complete model, it should satisfy the
electroweak gauge symmetries $SU(2)_L \times U(1)_Y$
with the left-handed neutrino $\nu_{\ell L}$
and its charged lepton counterpart $\ell_L$ in the
same $SU(2)_L$ doublet $(\nu_{\ell L}, \ell_L)^T$.
The flavor symmetry should be imposed on such doublet
to respect the gauge symmetries. Then, the flavor symmetry
has to be broken in order
to allow the left-handed neutrinos and charged leptons
to develop different mixing matrices $U_\nu$ and $U_\ell$,
respectively, such that the physical PMNS mixing
matrix $U_{\rm PMNS} \equiv U^\dagger_\ell U_\nu$
can be nontrivial \cite{Lam:2005va}.
Such flavor symmetry breaking
should happen at the same time as the electroweak
symmetry breaking. So if there is any flavor symmetry
that really dictates the neutrino mixing pattern,
it has to be the residual symmetry \cite{Lam:2007qc}
that survives the symmetry breaking processes.

% \gred{Para 4: Trend}
After symmetry breaking, a residual symmetry should
apply to the neutrino mass term whose diagonalization
gives the mixing pattern. It turns out a direct
connection can be established not just between the
symmetry transformation $G_\nu$
and the neutrino mass matrix $M_\nu$ but actually
between $G_\nu$ and the mixing matrix $U_\nu$
\cite{Lam:2006wm,Lam:2007qc,Lam:2008rs,Lam:2008sh}.
Then a correlation among the mixing parameters
(including three mixing angles and one leptonic
Dirac CP phase) can be established without involving
any model parameters or the neutrino mass eigenvalues
\cite{Dicus:2010yu,Ge:2011ih,Ge:2011qn}.
Study on such correlation was later further developed
under the name of {\it Sum Rule}.

% \gred{Para 5: Outline}
In this letter, we first summarize the major features
and unique predictions of the residual $\mathbb Z^s_2$
or $\overline{\mathbb Z}^s_2$ symmetries for Majorana
neutrinos. Since the
Dirac CP phase has not be fully established, we emphasize
its predicted values in terms of the already measured
three mixing angles \cite{Ge:2011ih,Ge:2011qn,Hanlon:2013ska}.
Especially, the JUNO experiment
provides very precise measurement of the solar angle
$\theta_s$ \cite{JUNO1st}. We show the current prediction of
$\delta_D$ has been significantly improved
and make an explicit comparison with the latest
measurement from the accelerator oscillation experiments
\cite{T2K:2025wet}.

\vspace{2mm}
\textbf{Residual Symmetries and CP Correlation with Mixing Angles}
--
As mentioned above, a residual symmetry is the one that survives
the electroweak symmetry breaking to allow different mixing
matrices for
neutrinos and charged leptons. Since neutrinos become
massive after symmetry breaking, the transformation
matrix $G_\nu$ of a residual symmetry directly applies to the mass
matrix $M_\nu$,
\begin{align}
  G^T_\nu M_\nu G_\nu
=
  M_\nu.
\label{eq:GMG}
\end{align}
For simplicity, we have chosen the mass basis of charged leptons
such that the PMNS matrix comes from the neutrino side alone.
Note that the above symmetry transformation invariance of
the mass matrix with a transposed $G^T_\nu$ applies for
Majorana neutrinos.

The direct connection between a symmetry transformation
matrix and the mass matrix is the essential feature of
residual symmetries. If a symmetry is really the one that
dictates the mixing pattern, it has to survive through
the mass generation process and directly restricts the
resulting mass matrix without interference from other
factors such as Yukawa couplings and the vacuum
expectation values.

The mass matrix $M_\nu$ of Majorana neutrinos
in the flavor basis can be diagonalized
as $U^T_\nu M_\nu U_\nu = D_\nu$ where
$D_\nu \equiv \mbox{diag}\{m_1, m_2, m_3\}$ is the diagonal
mass matrix in the mass basis, or equivalently,
$M_\nu = U^*_\nu D_\nu U^\dagger_\nu$.
Then one may replace the $M_\nu$ in \geqn{eq:GMG} to
obtain $G^T_\nu U^*_\nu D_\nu U^\dagger_\nu G_\nu
= U^*_\nu D_\nu U^\dagger_\nu$. Although the neutrino
mass eigenvalues $m_i$ are involved, it is possible to
obtain a mass indepenent solution,
\begin{align}
  U^\dagger_\nu G_\nu U_\nu
=
  d_\nu,
\quad \mbox{or} \quad
  G_\nu
=
  U_\nu d_\nu U^\dagger_\nu.
\label{eq:UGU}
\end{align}
Under the transformation of $d_\nu$,
the diagonal mass matrix $D_\nu$ should
also be invariant, $d^T_\nu D_\nu d_\nu = D_\nu$
which requires the $d_\nu$ to be diagonal with only
8 possibilities,
$d_\nu \equiv \mbox{diag}\{\pm 1, \pm 1, \pm 1\}$.
The diagonal transformation $d_\nu$
is actually the diagonal representation of the
same residual $\mathbb Z_2$ symmetries in the
mass basis.

It is interesting to observe that \geqn{eq:UGU}
is actually a direct connection between the residual
transformation matrix $G_\nu$ and the neutrino mixing
matrix $U_\nu$. As indicated by the first equation therein,
the mixing matrix $U_\nu$ can be obtained by diagonalizing
$G_\nu$. There is actually no need to first obtain the
mass matrix $M_\nu$ and then diagonalize it to determine
the mixing matrix $U_\nu$. Instead of such two step procedure,
the mixing pattern can be directly obtained from the
residual symmetry transformation. The mixing pattern
is really dictated by the residual symmetry without
involving any other factors. If we take the relation
in the other way around, the residual symmetry transformation
matrix $G_\nu$ can be reconstructed in terms of the mixing
matrix $U_\nu$ as explicitly shown by the second equation
in \geqn{eq:UGU}. This is a very neat and direct connection
between symmetry and observables.

Of the 8 possibilities of $d_\nu$ only two are independent,
$d^{(1)}_\nu \equiv \mbox{diag}\{-1, 1, 1\}$ and
$d^{(2)}_\nu \equiv \mbox{diag}\{ 1,-1, 1\}$.
Correspondingly, the two residual transformation matrices are,
\begin{subequations}
\begin{align}
  G_1(k)
& =
  \frac 1 {2 + k^2}
\left\lgroup
\begin{matrix}
  2 - k^2 & 2 k & 2 k \\
  & k^2 & - 2 \\
  & & k^2
\end{matrix}
\right\rgroup,
\\
  G_2(k)
& =
  \frac 1 {2 + k^2}
\left\lgroup
\begin{matrix}
  2 - k^2 & 2 k & 2 k \\
  & - 2 & k^2 \\
  & & - 2
\end{matrix}
\right\rgroup,
\end{align}
\end{subequations}
where $k$ is a free parameter. Both $G_1$ and $G_2$
are generators of a $\mathbb{Z}_2^s$ or
$\overline{\mathbb{Z}}_2^s$ symmetry, respectively.
Note that only one of these two $\mathbb Z_2$ symmetries
can exist as residual symmetry, especially after the
$\mu$-$\tau$ symmetry \cite{Xing:2015fdg,Xing:2022uax} that corresponds to
$d^{(3)}_\nu \equiv \mbox{diag}\{1, 1, -1\}$
and $G_3 = G_1 G_2$ is broken.

Although it seems like both $G_1(k)$ and $G_2(k)$
contain a model parameter $k$, they can predict unique
connection between the Dirac CP phase $\delta_D$ and 
three mixing angles, which are the atmospheric angle
$\theta_a\equiv \theta_{23}$, the solar angle $\theta_s$,
and the reactor angle $\theta_r$
\cite{Ge:2011ih,Ge:2011qn,Hanlon:2013ska}:
\begin{subequations}
\begin{align}
  \cos \delta_D 
& =
  \frac {(s^2_s-c^2_s s^2_r)(s^2_a -c^2_a)}
        {4c_a s_a c_s s_s s_r},
\\
  \cos \delta_D 
& =
  \frac {(s^2_s s_r^2-c^2_s)(s^2_a -c^2_a)}
        {4c_a s_a c_s s_s s_r},
\end{align}
\label{eq:sum_rule}
\end{subequations}
for $\mathbb{Z}_2^s$ and $\overline{\mathbb{Z}}_2^s$, respectively.
Here $s_{r,s,a}\equiv \sin\theta_{r,s,a}$ and
$c_{r,s,a} \equiv \cos\theta_{r,s,a}$ denote the sine and cosine functions
of the mixing angles. Since the $\mu$-$\tau$ symmetry
dictates a vanishing reactor angle ($\theta_r$ = 0)
and maximal atmospheric angle ($\theta_a = \pi /4$),
breaking it allows non-zero $\theta_a$ and $c^2_a - s^2_a$.
The ratio between these two deviations is correlated with
the leptonic Dirac CP phase $\delta_D$.

Although the transformation matrices $G_1(k)$ and
$G_2(k)$ are functions of a model parameter $k$,
the correlation above only involves physical
observables in neutrino experiments.
The property that a residual symmetry can establish
connection among physical observables has a close
analogy in the electroweak symmetry breaking.
Although the SM $SU(2)_L \times U(1)_Y$ gauge
symmetries are broken, the custodial symmetry
predicts a correlation among the gauge boson
masses ($M_Z$ and $M_W$) and the weak mixing
angle which are all physical observables. It
is readily possible to use physical observations
to justify such correlations. The residual
symmetry for the neutrino mixing pattern has
the same spirit as the custodial symmetry for
the gauge mixing \cite{Ge:2014mpa}. It is interesting to see
that both cases have the concept of mixing
pattern which is another similarity.

\vspace{2mm}
\textbf{CP Prediction with JUNO First Data}
--
The current global-fit result \cite{Esteban:2024eli}
has reached percentage level for the mixing angle measurement
and provided mild constraint on the Dirac CP phase.
In addition, the JUNO reactor neutrino experiment
just released their first data with better
measurement of  solar angle,
$s^2_s = 0.3092 \pm 0.0087$ \cite{JUNO1st}.
With this in mind, we would like to see how
the updated measurements affect the
CP phase prediction of the residual symmetry
sum rules. Especially, how the theoretical
predictions of the Dirac CP phase distribution
compare with its current measurement result.

The probability distribution function of $\cos \delta_D$
can be expressed by integrating over the mixing
angle distribution probabilities
\cite{Ge:2011qn,Hanlon:2013ska},
\begin{equation}
  \frac{dP(\cos\delta_D)}{d\cos\delta_D}
=
  \int \delta^p_D \, \mathbb{P}(s_r^2) \mathbb{P}(s_s^2) \mathbb{P}(s_a^2) \,
  d s_r^2 d s_s^2 d s_a^2,
\label{eq:cosdelta-distri}
\end{equation}
where $\delta^p_D\equiv \delta(\cos\delta_D-\bar c_D)$
is a $\delta$-function with $\bar c_D$ denoting
the RHS of \geqn{eq:sum_rule} to enforce the
residual symmetry sum rule. The probability
distribution function inside the integration
$\mathbb{P}(s^2_r, s^2_s, s^2_a) \equiv
\mathbb{P}(s_r^2) \mathbb{P}(s_s^2) \mathbb{P}(s_a^2)$
represents the prior probability distributions
of $(s^2_r, s^2_s, s^2_a)$ extracted from data
or global fits. Then, the probability distribution
of the Dirac CP phase $\delta_D$ can be obtained
with a simple transformation and Jacobian,
\begin{equation}
    \frac{dP(\delta_D)}{d\delta_D}
= 
  |\sin\delta_D| \frac{dP(\cos\delta_D)}{d\cos\delta_D}.
\label{eq:deltaD-distri}
\end{equation}
The prediction of the leptonic Dirac CP phase
$\delta_D$ would sensitively depend on the
input prior of mixing angles.
In the presence of measurement uncertainties of mixing angles,
the Dirac CP phase that predicted from the mixing angles is not
a fixed value but follows a distribution.
To calculate the predicted distribution of the Dirac CP phase from the
mixing angles, we take the current NuFIT result
\cite{Esteban:2024eli} with 
the best-fit values and the corresponding 
uncertainties of mixing angles.
% \gtab{tab:nufit} summarizes the inputs from the latest NuFit 6.0 
% updates \cite{Esteban:2024eli} in 2024 and the JUNO first
% data release \cite{JUNO1st}.

%\begin{table}[t]
%\centering
%\begin{tabular}{c|l|cc}
%\multicolumn{2}{c|}{} & NuFIT 6.0 (2024) & JUNO First Data \\
%\hline
%\multirow{3}{*}{NO} & $\sin^2\theta_{r}$ & $0.02195^{+0.00054}_{-0.00058}$ \\
%& $\sin^2\theta_{s}$ & $0.307^{+0.012}_{-0.011}$ & \gred{0.3092$\pm$ 0.0087} \\
%& $\sin^2\theta_{a}$ & $0.561^{+0.012}_{-0.015}$ \\
%\hline
%\multirow{3}{*}{IO} & $\sin^2\theta_{r}$ & $0.02224^{+0.00056}_{-0.00057}$ \\
%& $\sin^2\theta_{s}$ & $0.308^{+0.012}_{-0.011}$ &  \gred{0.3092$\pm$ 0.0087} \\
%& $\sin^2\theta_{a}$ & $0.562^{+0.012}_{-0.015}$ \\
%\end{tabular}
%\caption{Inputs of global-fit parameters from NuFIT 6.0
%\cite{Esteban:2024eli} (without SK atmospheric data) and
%the JUNO first release data \gred{\cite{JUNO1st}}.}
%\label{tab:nufit}
%\end{table}

As indicated by \geqn{eq:sum_rule}, the two deviations
($\theta_r$ and $c^2_a - s^2_a$) from the tribimaximal mixing
\cite{Harrison:2002er,Xing:2002sw,He:2003rm}
are proportional to each other. With
$|\cos \delta_D| \leq 1$ being limited from above,
the reactor angle $\theta_r$ needs to be nonzero
which has been firmly established by
Daya Bay \cite{DayaBay:2022orm}, RENO \cite{Shin:2020mue}, and 
Double-Chooz \cite{Soldin:2024fgt} with high precision.
Otherwise, neither deviations can happen.
A nonzero reactor angle $\theta_r$ is really
the key for going beyond the tribimaximal mixing.
With the precision measurements in the last 10
years, the uncertainty in $\sin^2 \theta_r$
has decreased from roughly 10\% to around 2.5\%.

For these three mixing angles, we take their
normalized distribution according to the
corresponding $\chi^2$ function, 
$\mathbb{P}(s^2_{a,r,s})\propto \exp(-\chi^2/2)$.
Note that the $\chi^2$ functions of $s^2_{r,s}$
are almost symmetric and follow a parabola. Then,
the probability distributions of $s^2_{r,s}$ can
be described by a normalized Gaussian distribution
function, $\mathbb P(s^2_{r,s}) \equiv
\exp[-(s^2_{r,s;{\rm BF}}-s^2_{r,s})^2)/2\sigma^2_{s^2_{r,s}}]/ \sqrt{2\pi}\sigma_{s^2_{r,s}}$. 
However, for the atmospheric angle, its distribution is 
asymmetric and has two local $\chi^2$ minima \cite{Esteban:2024eli}.
In this case, we calculate the $s^2_a$ distribution
probabilities using its exact $\chi^2$ function. 

\begin{figure}[t!]
\centering
\includegraphics[width=0.49\textwidth]{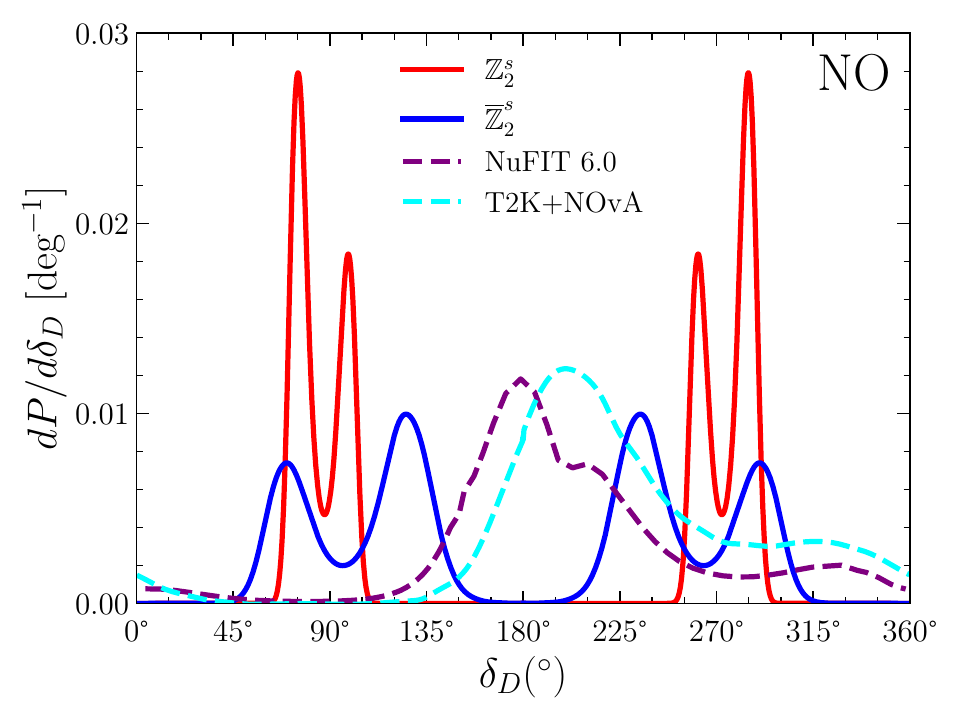}
\includegraphics[width=0.49\textwidth]{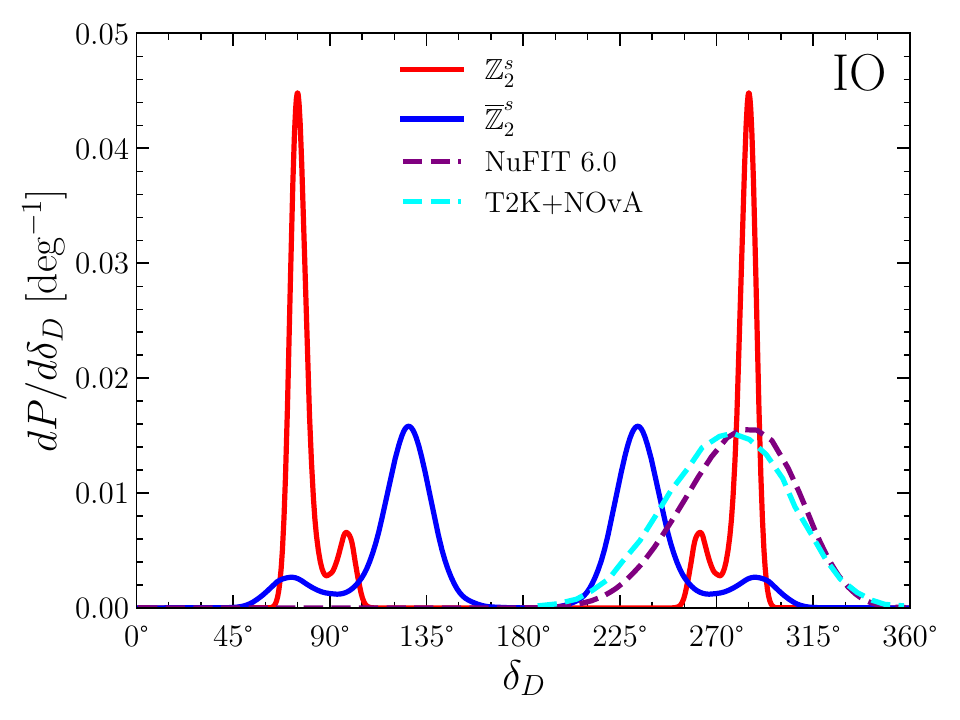}
\caption{The predicted probability
distributions of the leptonic Dirac CP phase
according to the residual $\mathbb Z^s_2$ (red)
or $\overline{\mathbb Z}^s_2$ (blue) symmetries
with inputs from the current
NuFIT 6.0 global fit plus the JUNO first data release.
In addition, the CP phase probability distributions
from the NuFIT 6.0 global fit (purple dashed) 
and the T2K plus NOvA joint analysis (cyan dashed)
are also shown for comparison. Both
normal (NO) and inverted (IO) orderings have been
shown in the upper and lower panels, respectively.}
\label{fig:deltaD-1D}
\end{figure}

\gfig{fig:deltaD-1D} shows the predicted probability
distribution function with the latest NuFit \cite{Esteban:2024eli} plus
the JUNO first data \cite{JUNO1st} (solid) for both $\mathbb Z^s_2$
(red) and $\overline{\mathbb Z}^s_2$ (blue).
In comparison with the previous results \cite{Hanlon:2013ska},
the predicted CP distribution shrinks quite significantly.
This is a manifestation of the fact that
the experimental uncertainties
on the mixing angles have improved quite a lot.

Comparing the solid curves that represent the
updated prediction of the $\delta_D$ distribution
with the previous results in 2013 \cite{Hanlon:2013ska},
the updated predictions have two local peaks.
Such a feature arises from the two
local minima in the $\chi^2$ distribution 
function of $s^2_a$.
To be more exact, the atmospheric angle $\theta_a$
has two local best-fit values in the lower
($\theta_a < \pi / 4$) or higher ($\theta_a > \pi / 4$) octants.
According to \geqn{eq:sum_rule}, switching octant
for $\theta_a$ would lead to a minus sign in the
predicted $\cos \theta_D$ and hence $\delta_D$
becomes $\pi - \delta_D$. This explains why
the two peaks mirror around the maximal CP point
$\delta_D = \pm \pi / 2$. However, such mirror
symmetry is not exact since the two octant solutions
of $\theta_a$ have different probabilities or weights \cite{Fogli:2012ua,Esteban:2024eli}.
For NO, the two peaks have comparable heights while
the IO case has a much larger disparity. But the
previous results in 2013 are dominated by a single
peak for the NO case, although the IO case has more
sizable second peak.
It seems that from the 2013 results
to the updated, the situation on the atmospheric angle
octant switches between NO and IO.

In addition, the major peak
predicted by $\mathbb Z^s_2$ is
below the maximal CP point, $\delta_D < \pi /2$ (or
$\delta_D > - \pi / 2$) for the updated results.
This corresponds to a positive $\cos \delta_D$ with
$s^2_a > c^2_a$ ($\theta_a > \pi / 4$) accordingly to
\geqn{eq:sum_rule} which is consistent with the preferred
higher octant best fit of $\theta_a$. The major peak moves above the
maximal CP point for the $\overline{\mathbb Z}^s_2$ case
since the first parenthesis in \geqn{eq:sum_rule}
switches sign between the two residual symmetries.
Moreover, the sum rule from the $\overline{\mathbb{Z}}_2^s$
residual symmetry gives a broader $\delta_D$ distribution than the 
$\mathbb{Z}_2^s$ case.

It is interesting to see that for all cases,
a vanishing CP effect ($\delta_D$ is either 0
or $\pi$) is not highly disfavored by the
two residual symmetries. Although the maximal CP
phase point ($\delta_D = \pm \pi / 2$) does
not have the highest probability since
$\theta_a = \pi / 4$ is not favored by data,
the two peaks are quite close to it.
This is a very important and distinct feature
that we can already conclude with the current data.
The residual $\mathbb Z^s_2$ and $\overline{\mathbb Z}^s_2$
symmetries prefer non-vanishing CP with probably
a sizable value. Let us take perturbative
expansion around the tribimaximal mixing
pattern, $\theta_r \equiv \delta_r$ is small and
$\theta_a \equiv \pi / 4 + \delta_a$
with both $\delta_r$ and $\delta_a$ denoting
the deviations. Up to the linear order,
\geqn{eq:sum_rule} becomes \cite{Ge:2011qn},
\begin{align}
  \mathbb Z^s_2:
  \cos \delta_D
\approx
  \frac {s_s}{c_s}
  \frac {\delta_a}{\delta_r},
\quad
  \overline{\mathbb Z}^s_2:
  \cos \delta_D
\approx
- \frac {c_s}{s_s}
  \frac {\delta_a}{\delta_r},
\label{eq:sum_rule_expanded}
\end{align}
The current best-fit gives
$\delta_r = 8.52^\circ$ ($8.57^\circ$) and
$\delta_a = 3.50^\circ$ ($3.56^\circ$)
for NO (IO) \cite{Esteban:2024eli}. For both mass orderings, the ratio
between the two deviations
$\delta_a / \delta_r \approx 0.4$ cannot be
compensated with the prefactor
$\tan \theta_s / 2 \approx 0.33$ for $\mathbb Z^s_2$
or
$\cot \theta_s / 2 \approx 0.75$ for $\overline{\mathbb Z}^s_2$
to make $\cos \delta_D$ close enough to 1.
A non-vanishing leptonic Dirac CP phase is guaranteed
by the residual $\mathbb Z^s_2$ and
$\overline{\mathbb Z}^s_2$ symmetries.

Besides the theoretical prediction of $\delta_D$ distribution 
from the sum rule, the experimental side also provides a $\delta_D$
distribution.
The first measurement of the Dirac CP phase was published by
T2K \cite{T2K:2019bcf} and NOvA \cite{NOvA:2019cyt} in 2019.
Very recently, the T2K and NOvA collaborations have also published
their combined analysis \cite{T2K:2025wet}, shown as cyan
dashed curves in \gfig{fig:deltaD-1D}, based on their
recent measurements \cite{T2K:2023smv, NOvA:2021nfi}.
Correspondingly, the global fit group updated the NuFIT 6.0
result \cite{Esteban:2024eli} in 2024 which is shown as
purple dashed curves.
The CP distributions are consistent between the global-fit result and 
T2K-NOvA joint analysis result in the case of IO, while for NO the CP distribution 
from the joint result is slightly shifted to the right-handed side.
As shown in the figure,
the best-fit values differ between the NO and IO cases.
While IO prefers a maximal CP phase $\delta_D \approx 280^\circ$,
NO prefers an almost vanishing one $\delta_D \approx 180^\circ$. 
Moreover, the NO has a broader distribution than the
IO case due to the existing tension between T2K and
NOvA results.

To quantify the preference of the theoretical
predictions of a model $M$ by a CP measurement data set $D$,
we take the Bayesian method \cite{Bayes} and 
define the following marginal likelihood,
\begin{align}
\hspace{-2mm}
  P(D|M) 
\equiv
  \int \mathbb P (D | f^M_\delta)
  \mathbb{P}(s^2_r) \mathbb{P}(s^2_s) \mathbb{P}(s^2_a) \, \mathrm{d} s^2_r \, \mathrm{d} s^2_s  \, \mathrm{d} s^2_a,
\label{eq:PDM}
\end{align}
where $f^M_\delta \equiv \delta_D (s^2_r, s^2_s, s^2_a)|_M$
is the theoretical prediction from the residual symmetry
as a function of the three mixing angles
and $\mathbb P (D | f^M_\delta)$ is the probability
distribution imposed by the data $D$. Comparing
with \geqn{eq:cosdelta-distri}, the $\delta^p_D$
function for imposing a fixed-value for the Dirac CP
phase which is a very special probability distribution
is replaced by a more realistic
$\mathbb P (D | f^M_\delta)$ with spread. If all the
probability distribution functions $\mathbb P$
extracted from data follow the Gaussian distribution,
after integration $P(D|M)$ is actually a manifestation
of the $\chi^2_{\rm min}$ that quantifies the
extent of fitting a data set $D$ with model $M$.
A larger value of $P(D|M)$, which corresponds to a
smaller $\chi^2_{\rm min}$, means better fitting.
To quantitatively compare
the preference of data between the two residual symmetries,
we adopt the Bayes factor,
$\text{BF} \equiv P(D|\mathbb{Z}_2^s) / P(D|\overline{\mathbb{Z}}_2^s)$.
To make it more explicit, a ${\rm BF} >1$
prefers $\mathbb Z_2^s$ and ${\rm BF} <1$
means that $\overline{\mathbb{Z}}_2^s$ has a better change. 
The two models have equal preference when ${\rm BF} =1$.

\begin{table}[t!]
\centering
\begin{tabular}{c|ccc}
\toprule
& \textbf{Model}  & \textbf{BF (1D)} & \textbf{BF (2D)} \\
\hline
\multirow{2}{*}{NO} & $\mathbb{Z}_2^s$ &  \multirow{2}{*}{0.42 (0.59)} &  \multirow{2}{*}{$0.30 $} \\
                    & $\overline{\mathbb{Z}}_2^s$  & \\
\hline
\multirow{2}{*}{IO} & $\mathbb{Z}_2^s$ & \multirow{2}{*}{2.11 (2.02)}  & \multirow{2}{*}{$2.40$}\\
                    & $\overline{\mathbb{Z}}_2^s$  & \\
\hline
\multirow{2}{*}{NO \& IO} & $\mathbb{Z}_2^s$  & \multirow{2}{*}{1.18 (1.13)} & \multirow{2}{*}{$1.35$}\\
                    & $\overline{\mathbb{Z}}_2^s$ & \\
\bottomrule
\end{tabular}
\caption{Bayesian evidence $P(D|M)$ and Bayes Factors BF
$\equiv P(D|\mathbb{Z}_2^s) / P(D|\overline{\mathbb{Z}}_2^s)$
with inputs from the NuFIT 6.0 (T2K-NOvA joint analysis)
and the JUNO first data. 
For the analysis with both orderings, we are assuming equal
probability for normal and inverted ordering.
}
\label{tab:evidence_results}
\end{table}

Our result of the Bayesian factor is summarized in
the BF (1D) column of \gtab{tab:evidence_results}
with inputs from the NuFIT 6.0 (T2K-NOvA joint analysis)
\cite{Esteban:2024eli} and the JUNO first data \cite{JUNO1st}.
For NO, the global-fit result prefers $\overline{\mathbb Z}_2^s$
which is consistent with \gfig{fig:deltaD-1D} where
the purple dashed curve has a larger overlap with the
blue solid curve for $\overline{\mathbb Z}_2^s$ than the
red solid one for $\mathbb Z_2^s$.
In the case of IO, 
the data prefers $\mathbb Z_2^s$ since the 
purple dashed curve overlaps more with red solid
curve than its blue solid counterpart.
It is also interesting to see that the peak 
position of the red solid curve is consistent with the purple dashed one, 
which means the theoretically predicted $\delta_D$ value from 
$\mathbb{Z}_2^s$ is very consistent with the
measurement. Note that the preference on the theoretical
models with input from the T2K-NOvA joint analysis is
slightly reduced than the global-fit case.

Moreover, since the global-fit
result has a light preference for NO, we also
present the combined Bayes factor,
\begin{equation}
  P(D|M)
\equiv 
  p_{\mathrm{NO}} \times P(D|M)_{\mathrm{NO}}
+ p_{\mathrm{IO}} \times P(D|M)_{\mathrm{IO}},
\end{equation}
to take into account the mass ordering weight factors
$p_{\mathrm{NO}}=0.574$ and $p_{\mathrm{IO}}=0.426$\footnote{ For the NuFIT 6.0 result, there is a $\Delta \chi^2=0.6$ preference for NO. This preference can be translated into priors $p_{\mathrm{NO}}$ and $p_{\mathrm{IO}}$. $\Delta \chi^2=\chi^2_\mathrm{NO}-\chi^2_\mathrm{IO}=0.6 \to \mathcal{L_\mathrm{NO}}/\mathcal{L_\mathrm{IO}}=\exp(-0.3)=0.741$. Fixing $\mathcal{L_\mathrm{NO}}=1$, the relative thickness $p_{\mathrm{NO}}=1/(1+0.741)=0.574$ and $p_{\mathrm{IO}}=0.741/(1+0.741)=0.426$.  }. 
The combined mass ordering analysis shows that the global-fit
result has only a mild preference for $\mathbb{Z}_2^s$. 
The results in the table show opposite preferences between 
the NO and IO cases. We expect the mass ordering to be resolved
with JUNO final result \cite{JUNO:2015zny}.

\begin{figure}[t]
\centering
\includegraphics[width=0.48\textwidth]{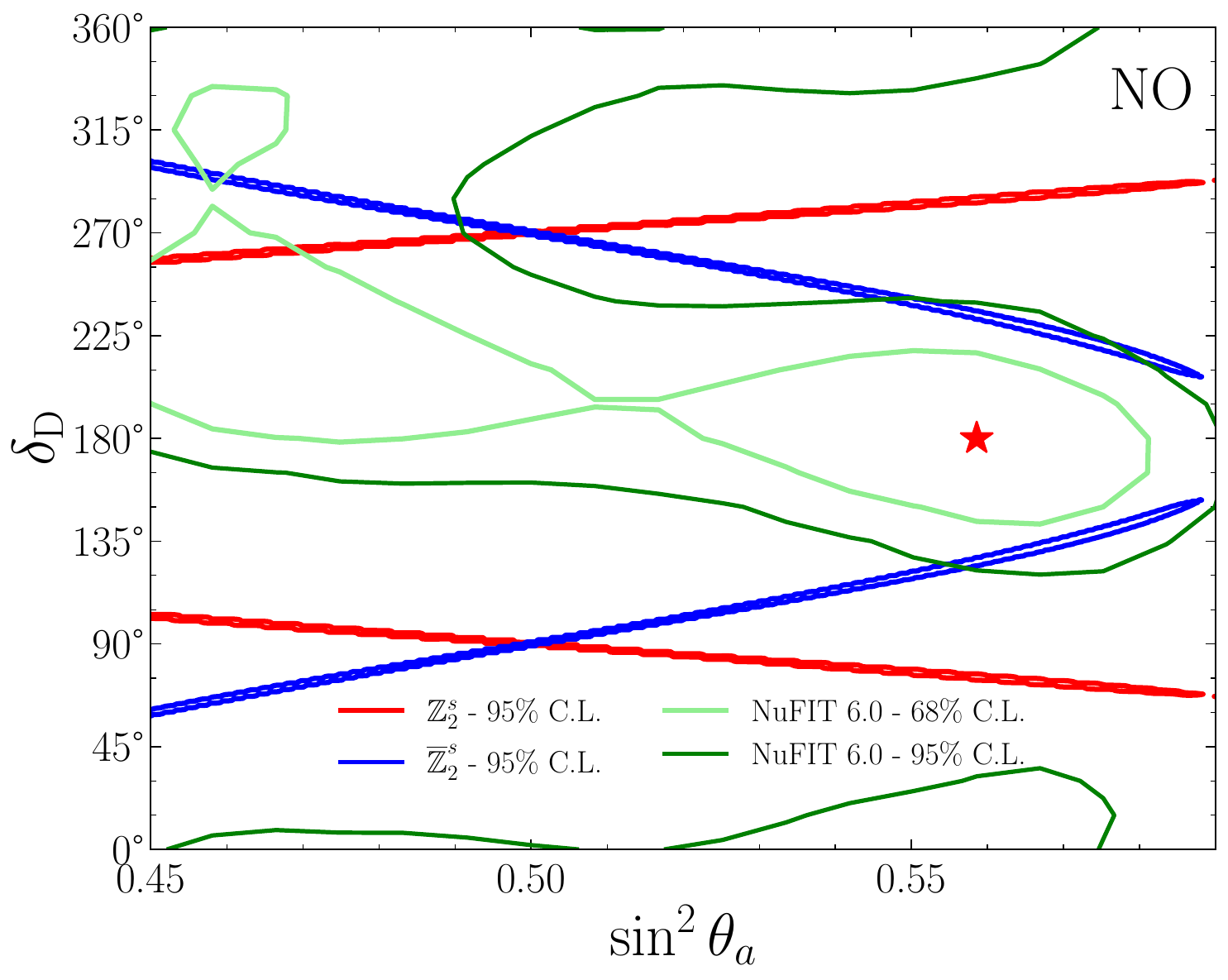}
\includegraphics[width=0.48\textwidth]{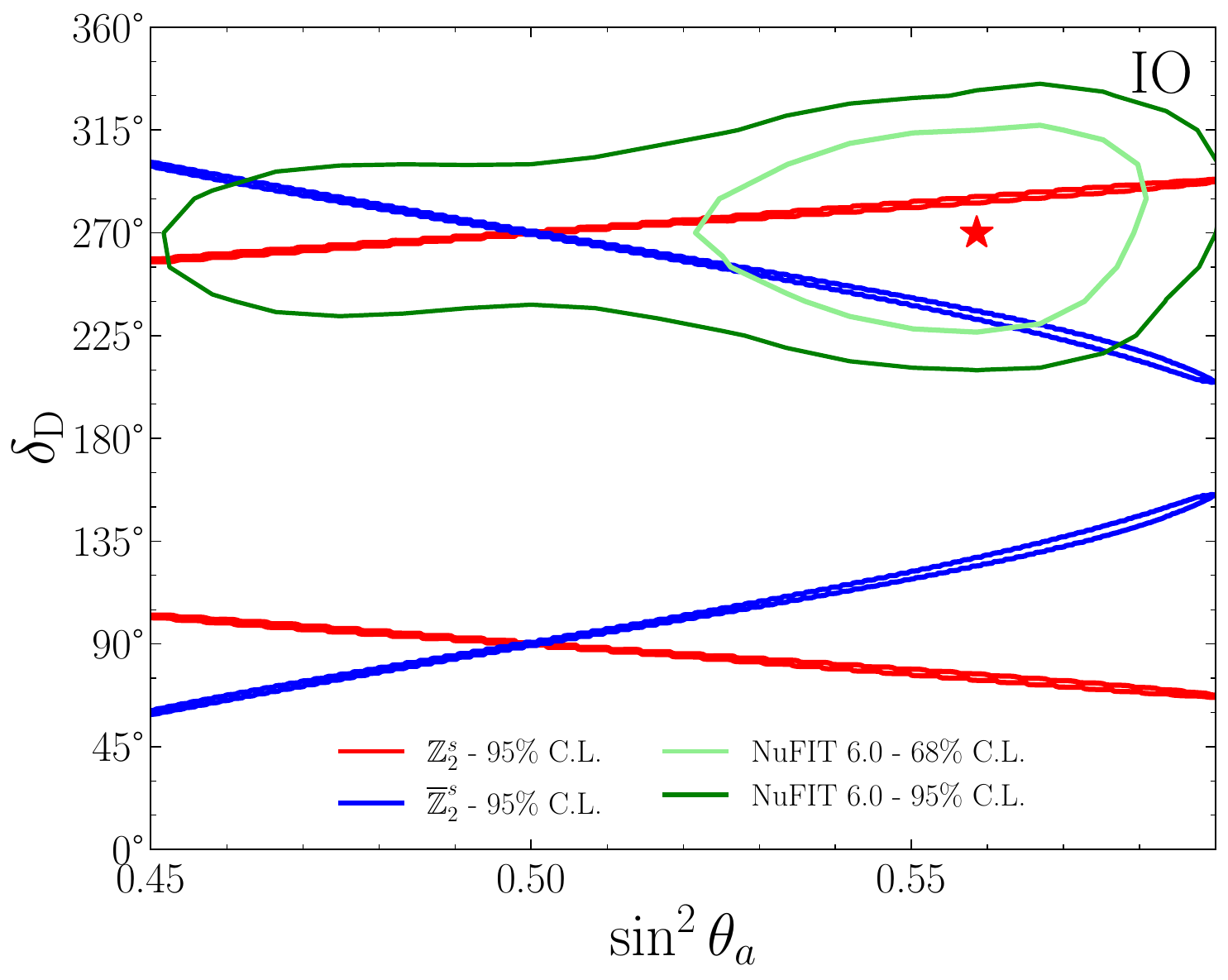}
\caption{Comparison between the theoretical prediction of the residual
$\mathbb Z^s_2$ (red) or $\overline{\mathbb Z}^s_2$ (blue) symmetries and the latest
NuFIT 6.0 results (green) for the two-dimensional correlated probability
distributions between the atmospheric angle $\theta_a$ and the
leptonic Dirac CP phase $\delta_D$. The upper panel is for NO
while the lower one is for IO.}
\label{fig:result_2D}
\end{figure}

\vspace{2mm}
\textbf{Correlation between Dirac CP Phase $\bf \boldsymbol \delta_D$ and Atmospheric Angle $\bf \boldsymbol \theta_a$}
-- 
Besides the predicted $\delta_D$ distribution from the sum rule 
as given in \geqn{eq:deltaD-distri}, it is interesting to 
consider the correlation behavior between $\delta_D$ and 
mixing angles. As reported in the current NuFIT result \cite{Esteban:2024eli},
the Dirac CP phase $\delta_D$ and the atmospheric angle $s^2_a$
are correlated, which can be seen in their two-dimensional
$\chi^2$ distribution, shown as light (68\% C.L.) and dark (95\% C.L.)
green contours in \gfig{fig:result_2D}.
The best-fit points are shown with a red star for both mass orderings. 
There exists a slightly negative correlation 
between $\delta_D$ and $s^2_a$ for NO and
a positive one for IO. 
Such information can also be useful
for testing the residual $\mathbb Z^s_2$ and
$\overline{\mathbb Z}^s_2$ symmetries.

To obtain the correlation from the sum rules,
we insert an additional
$\delta(s^2_a-s^2_{a,{\rm fix}})$ to fix $s^2_a$
in \geqn{eq:cosdelta-distri}.
Then only uncertainties of $s^2_r$ and $s^2_s$ are 
taken into account. For each fixed $s^2_a$, one may
obtain a 95\% C.L. range for the sum rule prediction
of $\delta_D$. Varying the fixed $s^2_a$, a
narrow band for the predicted $\delta_D$ as function
of the atmospheric angle $s^2_a$ can be formed
for $\mathbb{Z}_2^s$ (red) and $\overline{\mathbb{Z}}_2^s$
(blue), respectively.
One may see that the correlation behavior is opposite between 
$\mathbb{Z}_2^s$ and $\overline{\mathbb{Z}}_2^s$ predictions
which is independent on the mass ordering.
The predicted $\delta_D$ deviates from $\pi$ with
an increasing $s^2_a$ for $\mathbb Z^s_2$ (red) and
the opposite for $\overline{\mathbb Z}^s_2$ (blue).
Such difference in the correlation behavior arises
from the different factors in \geqn{eq:sum_rule},
which is $s^2_s-c_s^2 s^2_r$ for 
$\mathbb{Z}_2^s$ and $s^2_s s^2_r - c^2_s$ for 
$\overline{\mathbb{Z}}_2^s$. It becomes more explicit with
the sign difference between the two expanded forms
in \geqn{eq:sum_rule_expanded}.
Moreover, the best-fit values and the corresponding
uncertainties of $s^2_r$ and $s^2_s$
are almost the same between NO and IO.
As a result, the blue or red contours in \gfig{fig:result_2D}
are quite similar between the NO and IO cases. 

When comparing the theoretical prediction with the
global fit contours, it is interesting to see that
the $\mathbb{Z}^2_s$ symmetry has a positive
correlation between $\delta_D$ and $s^2_a$
for the $\delta_D > \pi$ branch which is
the consistent as the global-fit result for IO.
In addition, the $\overline{\mathbb Z}^s_2$ symmetry for $\delta_D > \pi$
has a positive correlation that is
consistent with the global-fit result for NO.

As mentioned above, we take the Bayesian factor to quantify the
global-fit result preference on the theoretical models. Here
we take the two-dimensional probability distribution of
the atmospheric angle $s^2_a$ and the leptonic Dirac CP phase
$\delta_D$ as data set $D$ to evaluate $P(D|M)$ in the similar
way as defined in \geqn{eq:PDM}. In such a way, the correlation
between $s^2_a$ and $\delta_D$ can be fully taken into account.
The results are 
summarized in the BF (2D) column of \gtab{tab:evidence_results}.
Taking this two-dimensional analysis into account, the preference 
of $\mathbb{Z}_2^s$ with NO decreases while the preference
of $\overline{\mathbb Z}_2^s$ increase in the IO case.
This is consistent with the contour plots in \gfig{fig:result_2D}
where $\overline{\mathbb Z}^s_2$ ($\mathbb Z^s_2$)
has negative (positive) correlation for NO (IO) which
are consistent with the NuFIT 6.0 contours. If both mass
orderings are taken into a combined analysis, the preference
for $\mathbb Z^s_2$ decreases.

\vspace{2mm}
\textbf{Conclusion}
--
The residual symmetry is by definition the one that can survive
symmetry breaking and apply to the neutrino mass matrix to
directly dictate the mixing pattern. A unique prediction of
the residual $\mathbb Z^s_2$ and $\overline{\mathbb Z}^s_2$
symmetries is a correlation among the three mixing angles and
the leptonic Dirac CP phase $\delta_D$ without involving
model parameters. Later named as sum rule, such correlation
involving only physical observables takes the same spirit
as the correlation between weak gauge boson masses and
the weak mixing angle that is predicted by the custodial
symmetry that also survives the weak symmetry breaking.

With the updated global fit and the JUNO first data release,
the CP prediction by the custodial symmetries has much
smaller uncertainty now. In particular, the vanishing
CP case ($\delta_D = 0$ or $\pi$) is theoretically
excluded now and the predicted CP distribution peaks
around the maximal CP value ($\delta_D = \pm \frac \pi 2$).
Since the experimental measurements have dependence on
the neutrino mass ordering, the residual $\mathbb Z^s_2$
and $\overline{\mathbb Z}^s_2$ symmetries are preferred by
IO and NO, respectively. The two-dimensional probability
distribution of $\delta_D$ and the atmospheric angle
$\theta_a$ with correlation can help distinguishing
these two residual symmetries. Moreover, breaking 
the $\theta_a$ octant degeneracy remains crucial for
giving more definite prediction of $\delta_D$
by eliminating the current double-peaked 
structure in the probability distributions to yield a 
single sharpe peak for each residual symmetry.

%%%%%%%%%%%%%%%%%%%%%%%%%%%%%%%%%%%%%%%%%%%%%%%%%%%%%%%%%%
\vspace{2mm}
\textbf{Acknowledgements}
--
The authors would like to thank Yue Meng for useful discussions.
The authors are supported by the National Natural Science
Foundation of China (12425506, 12375101, 12090060 and 12090064)
and the SJTU  First
Class start-up fund (WF220442604).
CFK is supported by IBS under the project code IBS-R018-D1.
SFG is also an affiliate member of Kavli IPMU, University of Tokyo.

\addcontentsline{toc}{section}{References}
\bibliographystyle{plain}

\end{document}